\documentclass[letter,twocolumn]{jpsj2}
\def\runauthor{T. Matsumura \textit{et al.}}

\title{
Spin Fluctuation and Crystal Field Excitation of a Heavy Fermion Compound YbAgGe studied by Inelastic Neutron Scattering
}

\author{
Takeshi \textsc{Matsumura}$^1$\thanks{E-mail address: tmatsu@iiyo.phys.tohoku.ac.jp}, 
Hideaki \textsc{Ishida}$^1$, Taku J. \textsc{Sato}$^2$, Kenichi \textsc{Katoh}$^3$, 
Yuzuru \textsc{Niide}$^3$ and Akira \textsc{Ochiai}$^4$
}

\inst{
$^1$Department of Physics, Graduate School of Science, Tohoku University, Sendai 980-8578\\
$^2$Institute for Solid State Physics, The University of Tokyo, Kashiwanoha, Kashiwa, Chiba 277-8581\\
$^3$Department of Applied Physics, National Defense Academy, Yokosuka 239-8686\\
$^4$Center for Low Temperature Science, Tohoku University, Sendai 980-8578
}

\recdate{\today}

\abst{
Inelastic neutron scattering experiment has been performed on a new heavy fermion compound YbAgGe  
for a polycrystalline sample at zero magnetic field. A quasielastic scattering 
and a crystal field excitation at 12 meV was observed. Both of them are broadened with an intrinsic width 
of 0.9 meV at 1.5 K which increases to about 3 meV at 300 K. 
The temperature dependence follows $\Gamma_0+A\sqrt{T}$ typically observed in Kondo systems.
Crystal field parameters were also deduced. To explain the result that only one excitation is observed at 12 meV, 
it is concluded that the $B_{22}$ term should be the main component. 
}

\kword{
heavy fermion, YbAgGe, spin fluctuation, crystal field, inelastic neutron scattering
}

\begin{document}
\maketitle

A ternary rare-earth compound YbAgGe has been attracting interest as a new Yb-based heavy-fermion 
material.~\cite{Katoh04,Umeo04, Budko04, Morosan04, Budko04b} 
One major aspect is the 
coexistence of Kondo effect and geometrical frustration in the \textit{kagom\'{e}}-lattice-like triangular coordination of Yb atoms 
in ZiNiAl-type crystal structure (hexagonal space group $P\bar{6}2m$). 
Heavy electron mass is observed which reaches about 400 mJ/mol K$^2$ at 2 K 
followed by successive magnetic phase transitions at quite low temperatures of 0.8 K and 0.65 K.~\cite{Katoh04,Umeo04} 
The two step transition is considered to be a characteristic of the geometrical frustration as observed in isostructural 
YbPtIn and YbRhSn.~\cite{Trovarelli00,Kaczorowski00} 
Another aspect of importance is the magnetic field induced non-Fermi-liquid (NFL) behavior which is typically manifested in the 
$-\ln T$ dependence of magnetic specific heat $C_{\text{mag}}/T$ in a wide temperature range from above 10 K 
to below 1 K.~\cite{Budko04} YbAgGe offers a good platform to study a crossover from magnetically ordered 
state with small moments to a Fermi-liquid (FL) state through a quantum critical point (QCP), as well as recently discovered NFL 
compound YbRh$_2$Si$_2$.~\cite{Trovarelli00b}

Characteristics of Kondo effect and heavy fermion are typically manifested in the magnetic specific heat.~\cite{Katoh04} 
At low temperatures $C_{\text{mag}}/T$ increases with decreasing temperature in proportion to $-\ln T$ from 
about 15 K to 3 K. Katoh \textit{et al.} analyzed the data with a single-ion Kondo model of $J=1/2$ for $T_{\text{K}}=24$ K. 
While the magnetic susceptibility at high temperatures follows Curie-Weiss behavior of well localized moment, 
$\chi(T)$ for $H\parallel a$ exhibits a weak maximum around 4 K, which suggests disappearance of the local moment. 
Although this is reminiscent of isostructural heavy fermion compound YbCuAl,~\cite{Murani85} YbAgGe is different in that 
a well defined crystal field (CF) anomaly is observed; $C_{\text{mag}}(T)$ exhibits a Schottky-type peak at around 
60 K, which indicates the first excited CF level at around 110 K.~\cite{Katoh04} 
Our final goal in YbAgGe will be to understand the whole process from the local-moment formation at zero field 
to the FL state at high fields through the QCP where the NFL state is observed. For this purpose, 
it is of essential importance to study correlation and fluctuation of the magnetic moments by inelastic neutron scattering (INS). 
The aim of the present study, as a starting subject, is to investigate spin fluctuation associated with the Kondo effect 
and to determine the CF level scheme, using polycrystalline sample at zero magnetic field.

INS experiment was performed using the LAM-D spectrometer at the KENS 
pulsed neutron source. LAM-D is an inverted geometry spectrometer using crystal analyzer mirrors.~\cite{Inoue93} 
The incident neutrons have a Maxwellian distribution of energy at room temperature and the scattered 
neutrons with an energy of 4.59 meV are selected by the (002) reflection of the pyrolytic graphite analyzer-crystals before 
being counted by the detector. The energy and momentum transfer are analyzed by the time of flight and 
the scattering angle. Two analyzer mirrors cover a range of low scattering angle around $\phi=35^{\circ}$; 
the other two cover high scattering angle around $\phi=85^{\circ}$ which we use to estimate the contribution 
from phonon scattering and to check the scattering-vector dependence of magnetic scattering. 

A collection of small pieces of polycrystals with total mass 14.2 g was sealed in an aluminium container 
with helium exchange gas and attached to a liquid-helium cryostat. The details of the sample preparation and 
characterization are described in Ref.~\citen{Katoh04}. 
The scattering intensity was put on an absolute scale in mb/meV/sr/Yb by comparing the intensity with that of the 
incoherent scattering from a vanadium standard sample. 
Contribution of phonon scattering was estimated from the data for $\phi=35^{\circ}$ and $\phi=85^{\circ}$ at 
high temperatures above 100 K; at low temperatures it was estimated from the high temperature results by 
calculating the Bose factor.

\begin{figure}[tb]
\begin{center}
\includegraphics[width=7.5cm]{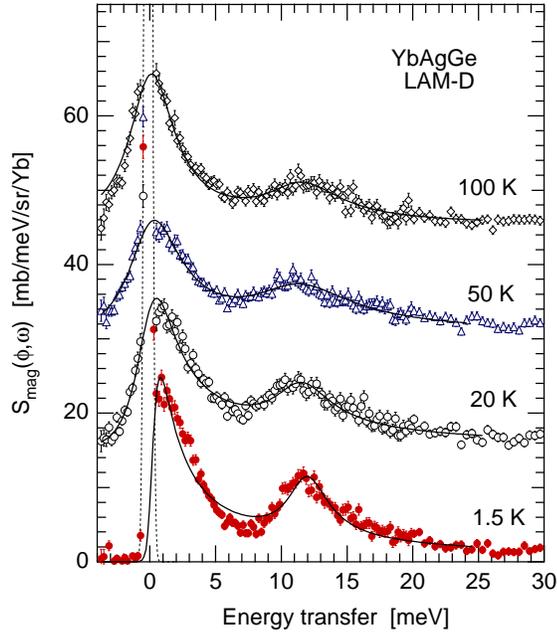}
\end{center}
\caption{Magnetic part of the neutron scattering function of YbAgGe per Yb ion. 
The data for 20 K, 50 K, and 100 K, are shifted by 15, 30, and 45, respectively. 
Solid lines are the fits with Lorentzian spectral functions. The dashed line is the measured spectrum of a 
vanadium standard sample normalized to the intensity of the incoherent scattering of the YbAgGe sample, 
representing the resolution function at 0 meV. 
}
\label{fig1}
\end{figure}

Figure \ref{fig1} shows the magnetic part of the neutron scattering function $S_{\text{mag}}(\phi,\omega)$ 
averaged for $\phi=35^{\circ}$ and $85^{\circ}$. 
Two features are clearly observed; one is a quasielastic (QE) scattering around $\hbar\omega=0$ meV and the other 
a magnetic excitation around 12 meV. Both of them are much broader than the energy resolution. 
The 12 meV excitation can be ascribed to the crystal field (CF) excitation of the Yb$^{3+}$ ions. 
A scenario that the first excited doublet is located at 12 meV above the ground doublet is consistent with 
the Schottky-type specific heat anomaly around 60 K reported in Ref.~\citen{Katoh04}.
No other significant peak structures were observed up to 60 meV.

We have analyzed the magnetic excitation spectra with  
\begin{equation}
 S_{\text{mag}}(\phi,\omega)=\frac{\hbar\omega}{1-\exp (-\hbar\omega/k_{\text{B}}T)}
 \sum_{i=1}^{2} f_{i} P_{i}(\hbar\omega; \Delta_i)\;,
\label{eq:1}
\end{equation}
where $P_i$ represents a spectral function of the $i$th excitation with an energy transfer $\hbar\omega=\Delta_i$ and 
$f_i$ the strength of the excitation. Lorentzian spectral function was assumed here. 
Fitting results are shown by the solid lines in Fig.~\ref{fig1}. 
With respect to the low energy excitation, the spectra above 20 K can be fitted well by the Lorentzian centered at 
$\hbar\omega=0$ meV, which indicates that the excitation is quasielastic. Although there remains a possibility that 
the excitation is inelastic with $\Delta$ less than 1 meV, it was not possible to determine within our energy resolution. 
The excitation at 12 meV can also be fitted well by the Lorentzian at all temperatures. 

The temperature dependences of the spectral strengths $f_i$ roughly exhibit the behavior of the bulk magnetic 
susceptibility. The strength of the QE scattering decreases with increasing temperature in accordance with 
the Curie term, while that of the 12 meV excitation do not vary with temperature very much 
since this corresponds to the Van-Vleck term. The decrease in the observed intensity in Fig.~\ref{fig1} is due to the 
temperature effect in eq.~(\ref{eq:1}). The sum of the two spectral strengths at 1.5 K becomes $\sim 160$ mb, 
which corresponds to the static susceptibility of 0.042 emu/mol and is consistent with the bulk susceptibility 
reported in Ref.~\citen{Katoh04}.

\begin{figure}[tb]
\begin{center}
\includegraphics[width=7.5cm]{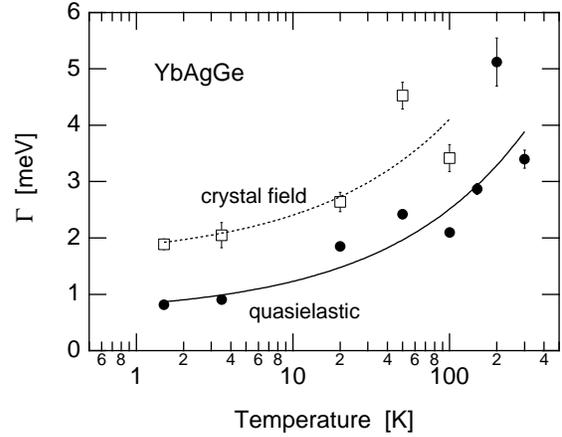}
\end{center}
\caption{Temperature dependence of the half width of the crystal field excitation and the quasielastic 
scattering. Solid and dashed lines represent fits with $\Gamma_{0}+A\sqrt{T}$. 
}
\label{fig2}
\end{figure}

Figure~\ref{fig2} shows the temperature dependence of the half width $\Gamma$ of the Lorentzian 
spectral functions. The widths of both QE scattering and CF excitation become broad 
with increasing temperature. Although the CF excitation has broader width than the 
QE peak, this is due to the broader energy resolution at 12 meV than at 0 meV; 
the two peaks have the same intrinsic widths. Even if we take into account the convolution 
effect by the resolution function, the intrinsic widths of the QE peak will be almost the same as 
those in Fig.~\ref{fig2}. 

Here, let us consider the CF level scheme. 
Since the eightfold multiplet of Yb$^{3+}$ ($J=7/2$) splits into four Kramers doublets, 
three excitations are expected at the lowest temperature. 
However, only one excitation was observed experimentally. It is necessary to look for a level scheme in which 
only the excitation from the ground doublet to the first excited doublet has a significant transition strength 
and the other excitations has negligibly small probability. 

Since the local symmetry of the Yb site is $C_{2v}$ when viewed from the $a$ axis, 
all the CF parameters $B_{kq}$ with even $k$ and $q$ up to six appear. 
Then, we first estimated the CF parameters from the electrostatic field of point charges in which we 
assumed negative charges for Ag and Ge, and a positive charge for Yb. As a result, we found that the $B_{22}$ term 
becomes the main source of the crystal field and the $B_{20}$ term modifies it slightly; higher order terms with 
$k=4$ and 6 are negligibly small. This is a straightforward result of the crystal structure of YbAgGe. 
Looking from the $c$ axis at the Yb ion at ($x$,0,0) on the $3f$ site and the surrounding Ag and Ge atoms, 
we notice that the Yb ion is in a strong $B_{22}$ type ($=x^2-y^2$ type) crystal field.~\cite{Gibson96} 
The $a$ axis becomes the principal axis of the CF anisotropy for this Yb ion. 
The anisotropy along the $c$ axis, which is a natural result of the hexagonal structure, induces the $B_{20}$ term. 
Since there is not an inversion symmetry, CF parameters of odd rank also exist; 
but these do not work as an actual crystal field, and the lowest order CF parameters are $B_{22}$ and $B_{20}$. 

Following the above observation, we have neglected the 4th and 6th order terms and calculated the CF eigenfunctions 
only with $B_{20}$ and $B_{22}$. First, with $B_{22}$ only, we found that only the transition strength between the ground and 
1st excited doublets has a significantly large value. Next, we modified the eigenfunctions by slightly including $B_{20}$ 
so that the magnetization curve and the temperature dependence of the magnetic susceptibility can be reproduced 
as well as possible. Our model of the CF parameters are $B_{22}=-14.77$ K and $B_{20}=1.477$ K, 
which give 1st, 2nd, and 3rd excited levels at 140 K ($=12$ meV), 228 K, and 323 K, respectively. 

The matrix elements of $J_{\perp}$, the perpendicular component of $\mib{J}$ to the scattering vector, 
for thus obtained eigenfunctions are listed in Table~\ref{tabl1}. 
It is noted that the direction of the principal axis of the CF anisotropy for the three Yb ions at ($x$,0,0), (0,$x$,0), and 
($-x$,$-x$,0), are different from each other by $120^{\circ}$. 
In addition, the matrix elements were averaged over the direction of the scattering vector to estimate 
the value for the polycrystalline sample. 
It can be seen from this table that the matrix element for the $(0,1)$ transition, 
which represents the transition between the ground state $|0\rangle$ and the 1st excited state 
$|1\rangle$, is much larger than those for the $(0,2)$ and $(0,3)$ transitions. 
Then, at low temperatures where only the ground state is populated, it is expected that there appear only one 
sizable excitation for $(0,1)$. When the temperature is elevated so that the 1st excited level is populated, the excitation 
for the $(1,2)$ transition is also expected.

\begin{table}[t]
\caption{Polycrystalline average of the squared matrix elements $|\langle i | J_{\perp} | j \rangle |^2$ for 
the crystal field eigenstates of Yb$^{3+}$ described in the text. }
\label{tabl1}
\begin{tabular}{lrrrr}
\hline
 & $|0\rangle$ & $|1\rangle$ & $|2\rangle$ & $|3\rangle$  \\
\hline
$\langle 0|$ & 20.17 & 3.99 & 0.11 & 0.02 \\
$\langle 1|$ &          & 9.61 & 6.70 & 0.45 \\
$\langle 2|$ &          &          & 7.32 & 4.63 \\
$\langle 3|$ &          &          &          & 15.13 \\
\hline
\end{tabular}
\end{table}

\begin{table}[t]
\caption{Polycrystalline average of the transition strengths of the $|i\rangle \rightarrow |j\rangle$ excitation 
at 1.5 K, 50 K, and 100 K. }
\label{tabl2}
\begin{tabular}{ccccccc}
\hline
$(i,j)$ & $(0,1)$ & $(0,2)$ & $(0,3)$ & $(1,2)$ & $(1,3)$ & $(2,3)$   \\
$E$ (meV) & 12.1 & 19.7 & 27.8 & 7.6 & 15.8 & 8.1 \\
\hline
$T$ (K) & \multicolumn{6}{c}{transition strength (mb)} \\
\hline
1.5 K & 31.2 & 0.51 & 0.055 & 0.00 & 0.00 & 0.00 \\
50 K & 27.3 & 0.47 & 0.051 & 3.89 & 0.15 & 0.44 \\
100 K & 17.0 & 0.33 & 0.038 & 8.6 & 0.40 & 2.40 \\
\hline
\end{tabular}
\end{table}

Calculated transition strengths at three different temperatures are listed in Table~\ref{tabl2} 
in unit of mb. Corresponding energy positions are also listed. 
The calculated result at 1.5 K is consistent with the experimental result that only one excitation is observed, 
and it can be concluded that the excitation corresponds to the $(0,1)$ transition. 
At elevated temperatures above 100 K, another excitation corresponding to the $(1,2)$ transition is expected 
to appear around 7.6 meV with its intensity about half as large as that of the $(0,1)$ transition. If this peak was 
observed experimentally, we could determine the energy level of the 2nd excited state; however, 
it was difficult to identify this excitation in the background of phonon scattering which dominates the 
observed intensity at high temperatures. More careful subtraction of phonon scattering using a reference sample 
is necessary to identify this peak.

\begin{figure}[tb]
\begin{center}
\includegraphics[width=7.5cm]{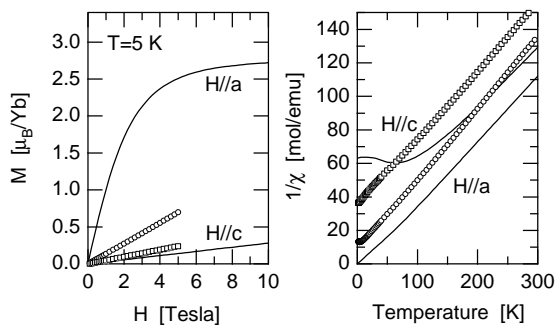}
\end{center}
\caption{Calculated magnetization curve at 5 K and temperature dependence of the inverse magnetic susceptibility 
for the crystal field model of YbAgGe described in the text. Experimental data in Ref.~\citen{Katoh04} is shown by the 
circles ($H\parallel a$) and squares ($H\parallel c$).
}
\label{fig3}
\end{figure}

The experimental results and the analysis of INS support the CF model in which the 
$B_{22}$ term is dominant. Figure~\ref{fig3} demonstrates the calculated magnetization and 
magnetic susceptibility for the CF parameters used in the analysis: 
$B_{22}=-14.77$ K and $B_{20}=1.477$ K. Molecular field is not considered; it just shifts $1/\chi(T)$ 
along the vertical axis and changes the scale of magnetic field in the $M(H)$ curves. 
Although the effective moment at high temperatures and the magnetic anisotropy along $a$ and $c$ axes 
are consistent with the experiment, $1/\chi(T)$ along the $c$ axis at low temperatures deviates from the experimental 
results. Since the calculated ground state has strong magnetic anisotropy within the $ab$ plane, $\chi(T)$ for $H\parallel c$ 
levels off below temperatures where the 1st excited state is depopulated. 
However, the experimental data follows the Curie-Weiss law down to the lowest temperature. 
Slight modification of $B_{20}$ cannot alter this behavior. Some mechanism which induce magnetic moments 
along the $c$ axis is necessary, which could be a hybridization with the conduction band or a ferromagnetic 
exchange in the Yb chain along the $c$ axis. 
Concerning $H\parallel a$, $1/\chi(T)$ in calculation is not far from the experimental results if we include a molecular field. Calculated magnetization exhibits a saturation moment of about 2.5 $\mu_{\text{B}}$, which also is roughly consistent 
with the experimental result that 2.0 $\mu_{\text{B}}$ is induced at 15 T.~\cite{Morosan04}

These CF parameters correspond to the set No. 12 in Table I of Ref.~\citen{Javorsky01} which can explain the 
bulk properties and the INS results of an isostructural ErNiAl.~\cite{Javorsky01} CF excitations in PrNiAl, NdNiAl, and ErCuAl, 
have also been measured, but the CF parameters have not been deduced.~\cite{Javorsky02}
It should be noted, in our analysis, that the sign of $B_{22}$ could not be determined since the calculated results 
little depend on the sign of $B_{22}$ in the present analysis. 

With respect to the spectral line shape, it is basically described by a Lorentzian in the whole temperature range 
of this study. The spectra in Fig.~\ref{fig1} may be regarded as single-site excitations. 
The QE line width follow roughly a $\Gamma_0+A\sqrt{T}$ law with intrinsic $\Gamma_0=0.9$ meV. 
The CF excitation is also broadened to the same width and exhibits the same temperature dependence. 
This behavior is commonly observed in Kondo systems in which the $4f$ electrons are almost 
localized; the width at $T=0$ indicates a single-site spin fluctuation and is regarded as a measure of 
$T_{\text{K}}$.~\cite{HollandMoritz94,Maekawa85} 
The value of $\Gamma_0=0.9$ meV and $\Gamma (300\text{K})\sim 3$ meV is slightly larger than 
the widths of YbBe$_{13}$, YbAuCu$_4$, YbPdCu$_4$, and YbPtBi, which are in the Kondo regime with 
well localized moments,~\cite{HollandMoritz94,Walter85,Severing90,Robinson95} 
but much smaller than those of YbCuAl and YbAgCu$_4$ in the intermediate valence regime,
~\cite{Murani85, HollandMoritz94,Severing90}
Then, from the aspect of spin fluctuation, YbAgGe can be categorized in the Kondo regime and has a relatively 
high characteristic temperature $T_{\text{K}}$. 

The transition temperatures of 0.8 K and 0.65 K in YbAgGe are several times 
lower than the N\'{e}el temperatures of other isostructural heavy-fermion Yb-compounds with lower $T_{\text{K}}$: 
$T_{\text{N}}=$3.4 K and 1.4 K in YbPtIn, 1.9 K and 1.7 K in YbRhSn, 3.5 K in YbPtSn, 
and 2.9 K in YbNiAl.~\cite{Trovarelli00,Kaczorowski00,Schank95,Diehl95} 
In view of the fact that the N\'{e}el temperatures of RAgGe for R=Tb-Yb roughly scale with the de Gennes factors,~\cite{Morosan04} 
it is considered that the RKKY interaction gives $T_{\text{N}}$ of a few Kelvin for the Yb compounds with ZiNiAl-type structure. 
Since a distinctive magnetic structure of frustrated moments has been observed in YbNiAl which orders at 2.9 K,
~\cite{Ehlers97} it is supposed that the low transition temperature in YbAgGe is due to the Kondo effect rather than the 
geometrical frustration.

At the lowest temperature of 1.5 K, the QE line shape deviates from the Lorentzian as can be seen in Fig.~\ref{fig1}. 
This could be ascribed to emergence of spin correlation leading to a scattering-vector dependence of the 
QE intensity as observed, e.g., in CeCu$_6$ and CeRu$_2$Si$_2$.~\cite{RossatMignod88,Appeli86} 
This point will be an important subject in future experiments using a single crystalline sample.

In conclusion, we have measured inelastic neutron scattering spectra of a new heavy fermion compound YbAgGe 
for a polycrystalline sample at zero magnetic field. The spectra exhibit a quasielastic scattering 
and a crystal field excitation at 12 meV, both of which are broadened because of the spin fluctuation associated 
with Kondo effect. The temperature dependence of the line width follows a typical behavior of $\Gamma_0+A\sqrt{T}$ 
observed in Kondo systems. The intrinsic width of 0.9 meV at 1.5 K and 3 meV at 300 K is relatively high in 
comparison with other Yb-based Kondo systems. 
Crystal field parameter was also deduced from the analysis of the observed excitations. 
Model calculation assuming the $B_{22}$ term as the main component successfully explains the excitation 
spectra. This is related with the ZiNiAl-type structure in which the Yb site is in a strong field of anisotropy.

\end{document}